\documentclass[%
 reprint,
 amsmath,amssymb,
 aps,
]{revtex4-2}

\usepackage{graphicx}
\usepackage{dcolumn}
\usepackage{bm}
\usepackage{braket}
\usepackage{jabbrv}
\usepackage{color}

\begin{document}

\preprint{APS/123-QED}

\title{Common Packing Patterns for Jammed Particles of Different Power Size Distributions}

\author{Daisuke Shimamoto}
\author{Miho Yanagisawa}%
 \email{myanagisawa@g.ecc.u-tokyo.ac.jp}
 \altaffiliation[Also at ]{Center for Complex Systems Biology, Universal Biology Institute, The University of Tokyo., Graduate School of Science, The University of Tokyo, Hongo 7-3-1, Bunkyo, Tokyo 113-0033, Japan}
\affiliation{%
Komaba Institute for Science, Graduate School of Arts and Sciences, The University of Tokyo, Komaba 3-8-1, Meguro, Tokyo 153-8902, Japan
}%

\date{\today}

\begin{abstract}

We introduce a model for particles that are extremely polydisperse in size compared to monodisperse and bidisperse systems. In two dimensions (2D), size polydispersity inhibits crystallization and increases packing fraction at jamming points. However, no packing pattern common to diverse polydisperse particles has been reported. We focused on polydisperse particles with a power size distribution $r^{-a}$ as a ubiquitous system that can be expected to be scale-invariant. We experimentally and numerically constructed 2D random packing for various polydisperse particles with different size exponents, $a$. Analysis of the packing pattern revealed a common contact number distribution for $a<3$ and a higher jamming point in $2<a<3$ than monodisperse systems. These findings demonstrate that the ambiguity of the characteristic length provides the common properties that leads to a novel classification scheme for polydisperse particles.

\end{abstract}

\maketitle

Polydisperse particles are omnipresent. Thermal particles, such as biomolecules in cells \cite{phillips2012physical,kiehn1970membrane} and athermal particles, such as cement \cite{taylor1909treatise} and gravel \cite{sammis1987kinematics,holtz1981introduction, grott2020macroporosity} are highly dispersed in size and shape. Regarding size polydispersity, fracture-produced particles such as impact-fractured objects 
\cite{oddershede1993self,ishii1992fragmentation,katsuragi2004crossover}, fault gouge under tensile stress \cite{sammis1987kinematics}, and rubble by the collision of rock \cite{grott2020macroporosity}, have power size distribution. 
In addition, critical phenomena result in a power size distribution of the clusters. It is known that correlation lengths related to the cluster size cutoff diverge at the critical point and that the cluster size follows a power size distribution \cite{tang1988critical,li2017sans}. It is also known that droplets with a power-law distribution arise because of self-organized criticality \cite{family1988scaling,stricker2022universality}. 
Thus, polydisperse particles with various power size distributions are ubiquitous and seemingly without any order; however, they have one thing in common: they have no apparent characteristic length.

In the context of jamming and glass transitions, such particle size dispersion has been considered based on monodisperse systems  \cite{bernal1960packing,o2003jamming,van2009jamming,karayiannis2009structure,liu2001jamming,charbonneau2011glass,charbonneau2012universal}. Previous studies have shown that size polydispersity suppresses crystallization at high packing fractions in two dimensions. Examples include bidisperse systems (size distribution with two peaks) with a size ratio of approximately 1.4  \cite{o2003jamming,kob1995testing,o2002random,zhang2009thermal,iikawa2016sensitivity} and systems with a slight size dispersity around the average size \cite{durian1995foam,takehara2014high,yanagisawa2021size}. In addition, more polydisperse bidisperse systems with larger size ratios than $1.4$
have been reported to exhibit unique phenomena, such as random packing at higher packing fractions \cite{yerazunis1965dense,zheng1995packing} and the appearance of multiple glass phases \cite{ikeda2021multiple,hara2021phase}.
Despite numerous experimental and numerical studies for polydisperse systems \cite{hermes2010jamming,durian1995foam,kwok2020apollonian,voivret2007space,hara2021phase,farr2009close,taylor1909treatise,o2003jamming,kob1995testing,o2002random,zhang2009thermal,iikawa2016sensitivity,yerazunis1965dense,zheng1995packing,ikeda2021multiple,hara2021phase}, there are no reports on common patterns to randomly packed polydisperse particles beyond the size polydisperse or on properties different from those of monodisperse particles. 

Because we anticipate that scale invariance results in common patterns of jammed particles beyond the size polydispersity, we have concentrated on systems with power size distributions with no apparent characteristic lengths, such as the mean and standard deviation of the sizes.

\begin{figure*}
\centering
\includegraphics[width=12.5cm, , bb = 0 0 321 159]{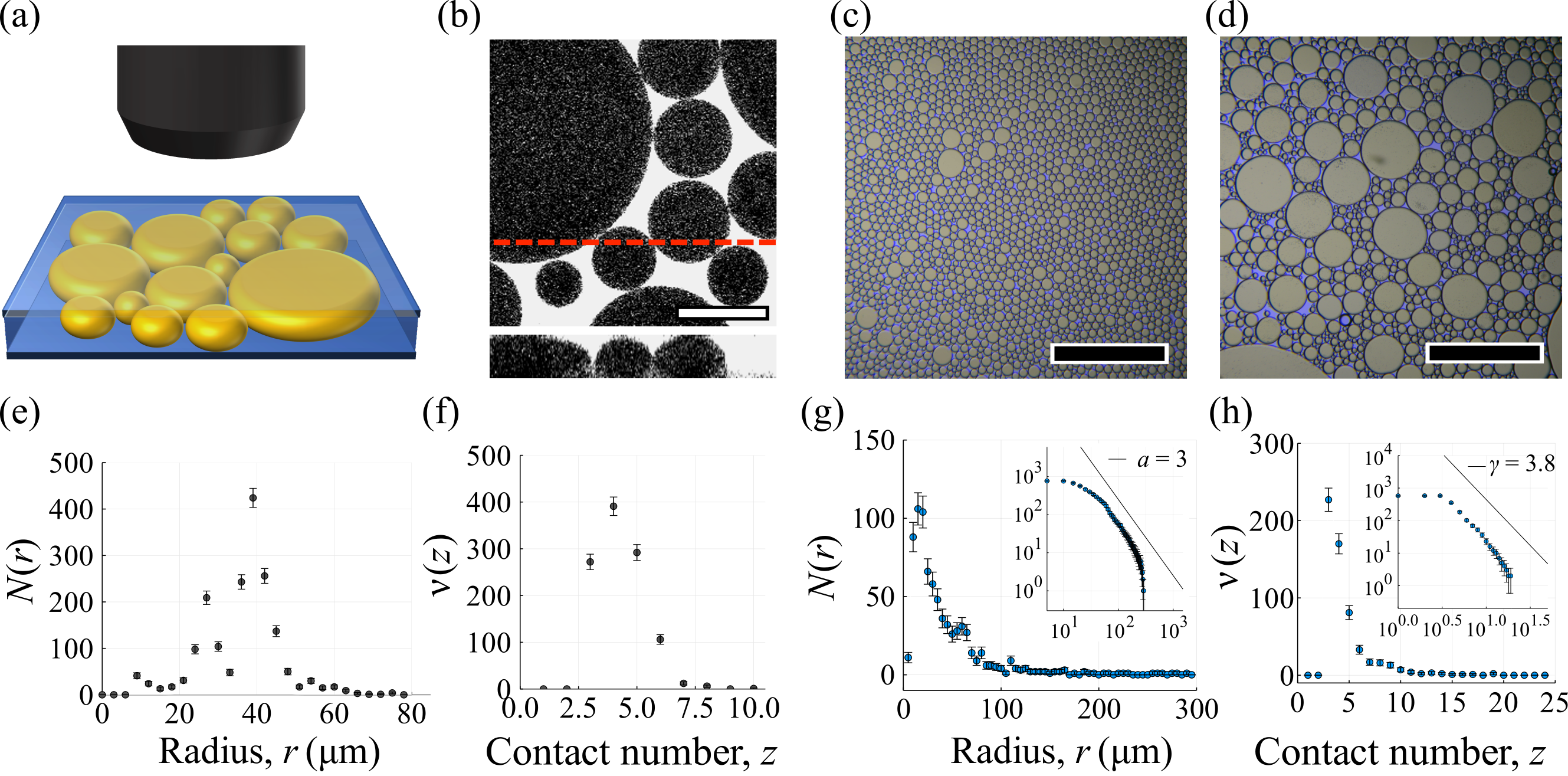}
\caption{(a) Schematic of the experimental setup of polydisperse droplets confined in 2D space. (b) Microscopic image of the polydisperse droplets from the top (top) and the cross-sectional image along the dashed line (bottom). Scale bar is 100 $\mu$m. (c-h) Microscopic images (c, d), droplet radius distribution, $N(r)$ (e, g), and contact number distribution, $\nu(z)$ (f, h) for bidisperse system (c, e, f) and polydisperse system (d, g, h). Scale bars in (c, d) are 1 mm. Cumulative distributions of radius and contact number as represented by log-log graphs in (g) and (h), respectively; (g) $N(r)$ follows a power size distribution of $a\simeq3$ in the region of one order of magnitude.}
\label{fig1}
\end{figure*}

We explored the random packing patterns and the jamming transitions by numerically  and experimentally employing particles with various power size distributions, and we contrasted them with bidisperse systems. The power distribution of the particle radius has the minimum and maximum cutoffs at $r_{\rm min}$ and $r_{\rm max}$. 
Although these $r_{\rm min}$ and $r_{\rm max}$ length scales exist due to constraints for physical realization, such as the finite number of particles and area, we have derived conditions under which these length scale effects can be neglected using experiments and simulations.
Therefore, this study contributes to describing the actual power size distribution, which can lead to practical scale invariance in jamming transitions.

\paragraph*{Methods}
Polydisperse particles with power size distribution were prepared by impact fracture of oil droplets in water. To the microtube containing 500 $\mu$L aqueous solution, 150 $\mu$L of mineral oil was added in 3 portions. Each time the oil was added, the microtube was tapped with a finger. Alternating the oil addition and tapping allows for the preparation of droplets with a power size distribution. The radii of prepared droplets are ranged from $14$ to $421$ $\mu$m.
Two-dimensional (2D) particles were prepared by sandwiching oil droplets in water between two glass plates having a thickness comparable to the diameter of the smallest particle (Figure 1(a)). This confinement deforms the particles into a pancake shape and prevents larger particles overhanging onto smaller ones (See also Supplementary Material). For the confirmation, we added a fluorescent molecule to the continuous phase to provide a clear distinction between the inside and outside of the particles. The pancake shape of the particles and the realization of a 2D system were confirmed by three-dimensional images with a confocal fluorescent microscope (Figure 1(b)).

In the numerical calculations, 4,000 circular soft-core repulsive particles were randomly placed in a square space with periodic boundary conditions as the initial condition. The radii of the particles $r$ are randomly set according to a given power size distribution $r^{-a}$. 
Typically, the size ratio $r_{\rm max}/r_{\rm min}$ was set to $10^2$. 
Increasing $r$ of all particles in the same ratio raises the packing fraction, and the particles move to the local minimum of the potential energy. Immediately after the total potential energy began to rise to a finite value, the particles immediately stopped expanding and relaxed further (see Supplementary Material).

\paragraph*{Common pattern of packed particles}
We analyzed the randomly packed particles with power size distribution $r^{-a}$ in 2D at the jamming point and compared with bidisperse system. Figure \ref{fig1}(c) shows a microscopic image of the packing pattern of bidisperse droplets. The packing fraction was $0.85\pm0.01$. To avoid crystallization, two differently sized particles were mixed. The ratio of the radii $r$ was approximately 1:1.5. From this image, we calculated the distribution of the particle radius $N(r)$ and distribution of the contact number $\nu(z)$ as shown in Figures \ref{fig1}(e) and (f). During the analysis of the contact number $z$, particles with $z\leq 2$ (not included in the contact network) were successively removed as rattlers. The contact number distribution $\nu(z)$ has one peak, which seems to be Gaussian suggested by previous work in bidisperse particles \cite{charbonneau2015jamming}. The average value is $\Braket{z}\simeq4.3$, which is similar to the ideal value $\Braket{z}$=4. 

\begin{figure*}
\centering
\includegraphics[width=15.4cm, bb = 0 0 323 156]{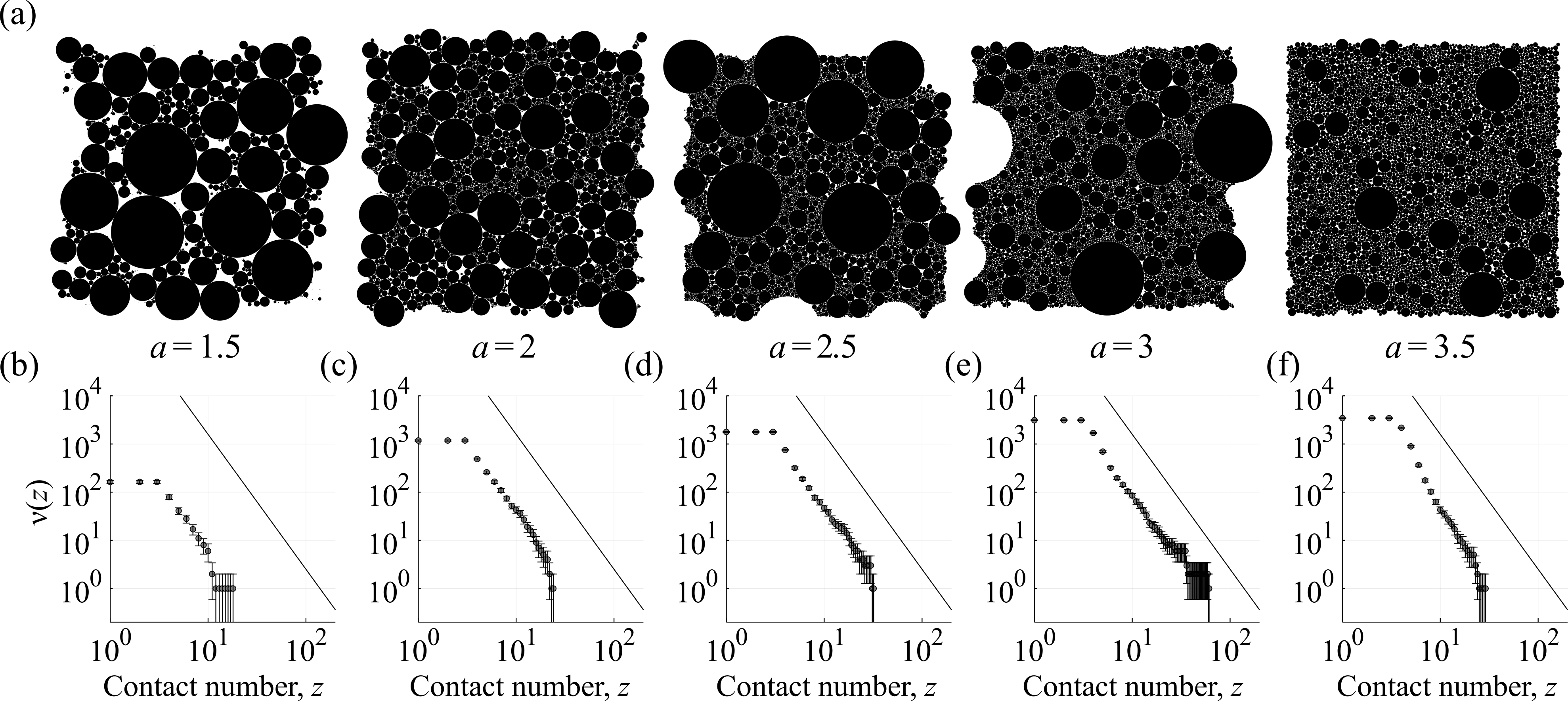}
\caption{(a) Examples of numerically produced packing patterns for various polydisperse systems $r^{-a}$, and (b)-(f) corresponding cumulative contact number distribution $\nu(z)$. From left to right, $a=$1.5, 2, 2.5, 3, 3.5.}
\label{fig2}
\end{figure*}

Similarly, the distributions of $N(r)$ and $\nu(z)$ for polydisperse particles were calculated from the microscopic images (Figure \ref{fig1}(d)), where the packing fraction was $0.94\pm0.02$. For this polydisperse system, $N(r)$ has a power size distribution of $a\simeq3$ in the region of one order of magnitude, as shown in Figure \ref{fig1}(g). The range of the power size distribution $r_{\rm max}/r_{\rm min}$ is approximately 10, where $r_{\rm min}$ and $r_{\rm max}$ are the minimum and maximum cutoff lengths, respectively. 
The resulting average contact number was $\Braket{z}\simeq4.5$ (Figure \ref{fig1}(h)). 
On the other hand, the contact number distribution $\nu(z)$ also follows a power distribution $\nu(z)\propto z^{-\gamma}$, and not a Gaussian, but decayed more rapidly than $N(r)$ (See insets of Figures \ref{fig1}(g)(h)). The exponent of $\nu(z)$ was found to be $\gamma\simeq3.8$.

To investigate the generality of the contact number distribution for polydisperse systems, $\nu(z)\propto z^{-\gamma}$ with $\gamma\simeq3.8$ suggested by the experiments, we numerically produced randomly packed patterns of various polydisperse particles with different $a$ values. Figure \ref{fig2}(a) shows examples of numerically produced packing patterns for $a$=1.5, 2, 2.5, 3, and 3.5.
For each $a$, the contact number distribution $\nu(z)$ was calculated, as plotted in Figures \ref{fig2}(b)-(f). Note that $\nu(z)=0$ for $z\leq 2$, independent of $a$, because the rattlers were removed. Figures \ref{fig2}(b)-(e) show that $\nu(z)$ for $a<3$ follows a power-law distribution with $\gamma\simeq3.8$ independent of $a$. This exponent agrees with the experimentally suggested value of $\gamma\simeq3.8$ (Figure \ref{fig1}(h)). In addition, we have confirmed that $\Braket{z}$ is approximately $4$ regardless of $a$, same as that of the bidisperse system. These results demonstrate a common property for the contact number distribution of randomly packed polydisperse particles $a<3$.
For larger exponent ($a>3$), deviation from the power law was observed.

To explain the contact number distributions of $\nu(z)\propto z^{-\gamma}$ with constant $\gamma\simeq3.8$ in $a<3$, we modeled the system with dimension $d$ with two assumptions: (i) If the particle size follows a power distribution, then the contact number also follows a power distribution $\nu(z)\propto z^{-\gamma}$ for $z\geq d+1$ and $\nu(z)=0$ for $z\leq d$, for all particles except rattlers are in contact with at least $d+1$ particles. In general, $\nu(z)$ cannot be estimated from $r(z)$, but here, we assume a power distribution.
(ii) $\Braket{z}=2d$. This assumption is based on the fact that $\Braket{z}$ is $2d$ at the jamming point, and easily derived from Laman's theorem \cite{laman1970graphs} when $d=2$.

From the assumption (ii), the following equality holds:
\begin{eqnarray}
\Braket{z} =\frac{
    {\sum_{z=d+1} z\nu(z)}
}{
    \sum_{z=d+1}\nu(z)
}
=2d.
\end{eqnarray}
For $d=2$, this can be expressed as follows using the assumption (i):
\begin{eqnarray}
\frac{
    \zeta{\left(\gamma-1, d+1\right)}
}{
    \zeta{\left(\gamma, d+1\right)}
}
=4,
\label{zeta}
\end{eqnarray}
where $\zeta(z, a)$ denotes the Hurwitz zeta function.
Solving Eq. \ref{zeta} yields $\gamma= 3.83\ldots$ when $d=2$, which explains  the property obtained experimentally and numerically (Figures \ref{fig1},\ref{fig2}). The exponent $\gamma\simeq 3.8$ is considered sufficient as long as assumption (i) remains valid. It  suggests that the assumptions is not satisfied with $a>3$ (Figure \ref{fig2}(f)).

\paragraph*{Classification of polydisperse systems from jamming point}

\begin{figure}
\centering
\includegraphics[width=86mm, bb = 0 0 480 258]{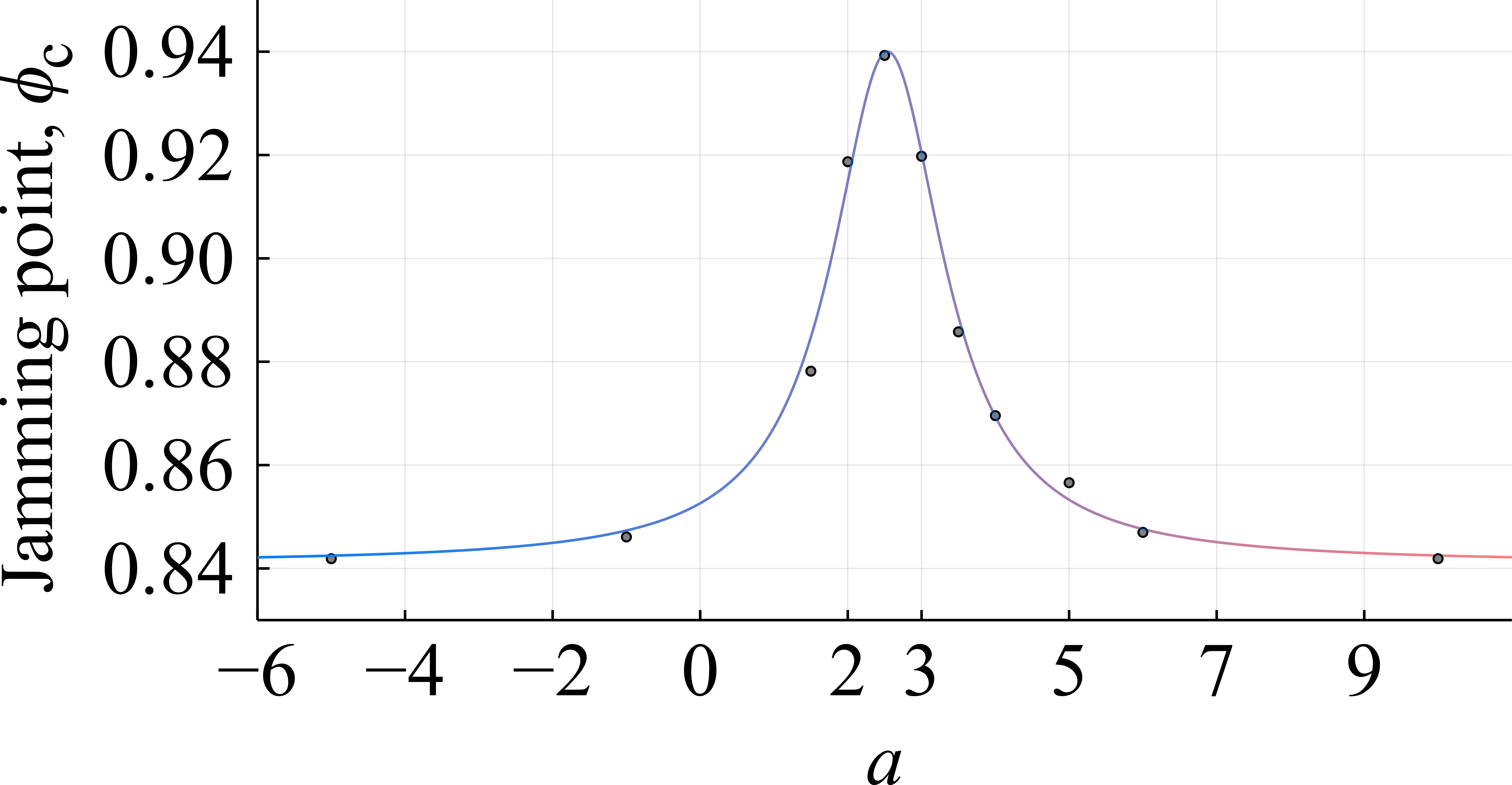}
\caption{(a) Dependece of pressure $P$ on the packing fraction $\phi$ for various $a$. For better visibility, the curves were shifted vertically. (b) Dependence of $\phi_{\rm c}$ on $a$.}
\label{fig3}
\end{figure}

We identified that the common contact number distribution $\nu(z)\propto z^{-3.8}$ holds for $a<3$. To clarify the physical meaning of the range $a<3$ and classification of such polydisperse systems, we investigated the jamming transitions for various polydisperse systems. 
Numerically obtained jamming point $\phi_{\rm c}$ was plotted against $a$ in Figure \ref{fig3}.
For extremely small and large values of $a=-5$ and $a=10$, respectively, $\phi_{\rm c}$ is close to the well-known value for bidisperse or small dispersity systems $\sim0.84$ \cite{o2002random,takehara2014high}. However, $\phi_{\rm 
c}$ reaches a maximum value within the range $2<a<3$. This means that the exponent range $2<a<3$ exhibits particularly strong characteristics of polydisperse systems.

\paragraph*{Characteristic length in packing}
We have shown that the 
The contact number distributions have a common exponent $3.8$ for $a<3$ and a significantly higher packing fraction than the bidisperse system for $2<a<3$. 
To explain the reason, here we discuss the implications of this range in terms of characteristic length scales: we examine the effect of the upper and lower limits of the particle size, i.e., $r_{\rm max}$ and $r_{\rm min}$ on the packing pattern.

First, we consider the effect of $r_{\rm min}$, based on a comparison with an ideal system with $r_{\rm min}=0$.
For $a<3$, the total area of particles under $r_{\rm min}$, $\int_0^{r_{\rm min}}\pi r^2N(r)dr$, can be made as small as desired by taking $r_{\rm min}$ sufficiently small, which renders the effect of $r_{\rm min}$ negligible. However, for $a>3$, the divergence of $\int_0^{r_{\rm max}}\pi r^2N(r)dr$ makes the packing of particles with an ideal distribution undefinable. For the actual system, we must set a finite $r_{\rm min}$ because the limit of $r_{\rm min}\rightarrow0$ cannot be taken. It means that $r_{\rm min}$ remains as the characteristic length for $a>3$.

Next, we consider the effect of $r_{\rm max}$ based on a comparison with the complete packing, which is a packing without voids constructed by optimal arrangement.
In a two-dimensional system, the minor numbers of small particles make complete packing impossible for $a<2.3\ldots$. 
The condition of $a\simeq 2.3$ corresponds to Apollonian packing, which is a complete packing with the smallest $a$ ($=d_{\rm A}+1$, where $d_{\rm A}$ is a fractal dimension of Apollonian packing) and the smallest number of particles \cite{manna1991precise, kwok2020apollonian}.
Similarly, when $a$ is too small for the random packing, the space around large particles is not sufficiently filled. 
The random packing contains voids in comparison to the optimally ordered Apollonian packing, yet it nevertheless achieves a packing fraction of 0.93, which is noticeably greater than the value of bidisperse systems $\sim$0.84. Due to the scarcity of small particles, large particles contact each other to form a pattern. Consequently, $r_{\rm max}$ appears as the characteristic length.

Thus, for sufficiently large $a>3$, a characteristic lengths $r_{\rm min}$ emerge, and the scale-free nature is broken, which violates assumption (i). This is analogous to the absence of small length cutoffs in fractal figures.
Furthermore, both $r_{\rm min}$ and $r_{\rm max}$ have negligible effects on the pattern in the range $2<a<3$. 
This ambiguity in the characteristic length scale enhances the polydispersity and leads to a high packing fraction for polydisperse systems.

We have demonstrated that the packing of polydisperse particles with $N(r)\propto r^{-a}$ has a common property for the contact number distribution when the exponent $a$ is smaller than $3$. Furthermore, the power distribution with $2<a<3$ corresponds to the range in which a particularly strong polydispersity appears during the jamming transition.
The proposed classification based on $a$ may be applicable to various polydisperse systems with a power size distribution \cite{meibom1996composite, grott2020macroporosity} and general probability distributions by generalizing $a$ as follows: 
\begin{equation}
    a=-\frac{\ln{N(r)}}{\ln{r}},
\end{equation}
where $r\ll 1$.

Finally we discuss how our findings contribute to the understanding and application of polydisperse systems. Scale-free nature of a polydisperse system may allow us to derive exact solution of the jamming point $\phi_{\rm c}$ at $r_{\rm min}\rightarrow 0$. In addition, when such polydisperse systems are at very high packing fraction beyond the jamming point, another common property may be derived from the analysis of particle dynamics and shape deformation. Furthermore, the condition $d_{\rm A}+1<a<d+1$, where the scale-free nature appears, will enhance the distinctive characteristics of polydisperse particles for general dimensions.

\begin{acknowledgments}
This research was funded by the Japan Society for the Promotion of Science (JSPS) KAKENHI (grant numbers 21H05871, and 21K18596) and Japan Science and Technology Agency (JST) program FOREST (grant number JPMJFR213Y). The authors thank Dr. Shio Inagaki (Kyushu University), Dr. Kyohei Takae (The University of Tokyo), Dr. Norihiro Oyama (Toyota Central R\&D Labs., Inc.), and Mr. Yusuke Hara (The University of Tokyo).
\end{acknowledgments}


\bibliographystyle{apsrev4-2}
\bibliography{poly_abrv2}

\end{document}


\preprint{APS/123-QED}

\title{Supplementary Material for "Common Packing Patterns for Jammed Particles of Different Power Size Distributions"}

\author{Daisuke Shimamoto}
\author{Miho Yanagisawa}%
 \email{myanagisawa@g.ecc.u-tokyo.ac.jp}
 \altaffiliation[Also at ]{Center for Complex Systems Biology, Universal Biology Institute, The University of Tokyo., Graduate School of Science, The University of Tokyo, Hongo 7-3-1, Bunkyo, Tokyo 113-0033, Japan}
\affiliation{%
Komaba Institute for Science, Graduate School of Arts and Sciences, The University of Tokyo, Komaba 3-8-1, Meguro, Tokyo 153-8902, Japan
}%

\date{\today}

\maketitle

Distilled water (UltraPure™ DNase/RNase-Free Distilled Water, Invitrogen, Waltham, MA, USA) with a surfactant Tween 20 (Sigma-Aldrich, St. Louis, MO, USA) was used as a continuous phase. Mineral oil (Nacalai Tesque, Kyoto, Japan) with a surfactant Span 80 (Tokyo Chemical Industry Co., Tokyo, Japan) was used as a dispersed phase. To facilitate identification of the oil-in-water (O/W) droplets, a lipophilic dye, capsanthin in vegetable oil (Tokyo Chemical Industry Co.) was added to the mineral oil.

We used O/W droplets to prepare randomly packed particles. An aqueous solution containing 1 wt\% Tween 20 was used for the continuous phase and a mineral oil containing 0.1 wt\% Span80 and 2 wt\% capsanthin oil for the dispersed phase.

For making monodisperse droplets, we used a centrifugal microfluidic device, which is a modified version of the previously reported device \cite{maeda2012controlled}. The device consists of three parts: a glass capillary, a micropipette tip (Labcon, Petaluma, CA, USA), and a microtube. The glass capillary with a thin tip of $\sim 30\,\mu$m and length of 8 mm from the tip was fabricated from the ready-made capillary (outer diameter: 1 mm, inner diameter: 0.6 mm; G-1, Narishige, Tokyo, Japan) by using a puller (PC-10, Narishige) and microforge (MF-900, Narishige). The capillary was attached to the end of a micropipette tip (200$\mu$L standard yellow pipette tip). 
The micropipette tip was filled with 80 $\mu$L of mineral oil and fixed on the microtube containing 500 $\mu$L of aqueous solution by passing through the 6 mm diameter hole drilled in the lid of the microtube.
The device was centrifuged at 6,000 rpm for 1 minute using a table-top centrifuge (Wako Pure Chemical Industries, Osaka, Japan). The prepared droplets have two different sizes, approximately $50\, \mu$m and $80\, \mu$m (See Figure 1(e)). 

To prepare the randomly packed particles in quasi-two-dimensional (quasi-2D) space, the O/W droplets were confined between two slide glasses (76 mm $\times$ 26 mm, thickness $\sim0.9$ mm, Matsunami, Osaka, Japan, S1111). These glasses were laminated together with $\sim$50$\,\mu$m-thick double-sided tape to nearly match the diameter of the smallest droplet $\sim28\,\mu$m. To cover the glass surface with an aqueous phase, the hydrophilicity of the glass slide surface was improved by using the plasma cleaner (Harrick Plasma, PDC-32G, Ithaca, NY), which reduces the friction with the oil droplets. This quasi-2D confinement deforms the droplets into a flat pancake shape and eliminates overlap between droplets. Hence, the center and edges of the droplet can be clearly identified.
The water surrounding the droplet evaporates very slowly over time ($\sim5\%/$h). This process increases the total area fraction occupied by the droplets (i.e., the packing fraction $\phi$), while maintaining the area of each droplet. Here we determined the moment at the first avalanche occurred as the jamming point and analyzed the packing pattern at that point.

\begin{figure}[tbhp]
\centering
\includegraphics[width=86mm, bb = 0 0 250 250]{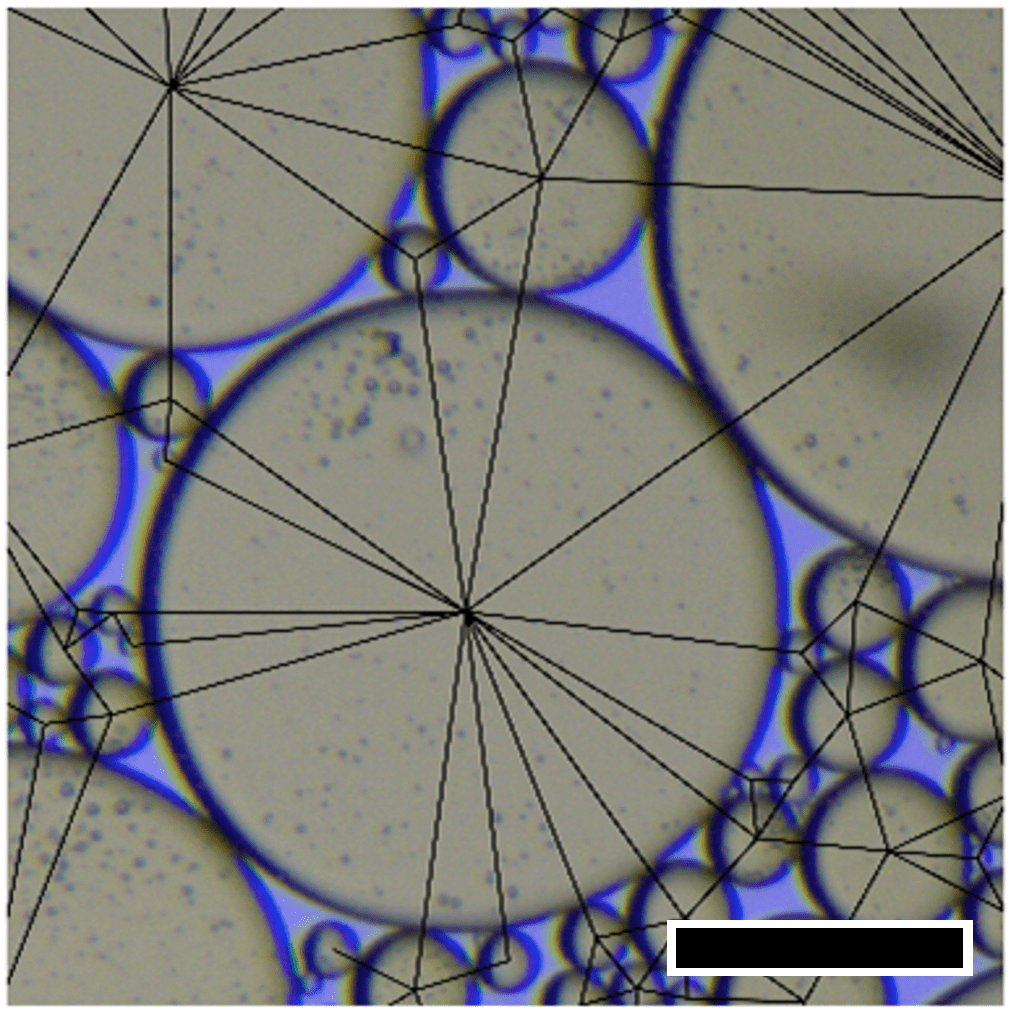}
\caption{Micrograph of the packed droplets with the automatically detected contact lines. Scale bar is 200 $\mu$m.}
\label{contact_detection}
\end{figure}

2D images of droplet packing were acquired using a camera (a2A5328-15ucPRO, Basler) attached to a microscope (SZX16, Olympus). The images were analyzed by using free NIH software, ImageJ (\url{http://rsb.info.nih.gov/ij/}). The droplets in the images were detected by binarization after removing noise with a median filter and FFT bandpass filter.
When the distance between the droplet surfaces was less than $1$ pixel, the droplets were determined to be in contact with each other (Supplementary Material, Figure S\ref{contact_detection}). The automatically detected contacts were confirmed on the microscope image and were well matched, with an error of approximately 4\%.
Fluorescent molecules, 10 $\mu$M TAMRA (Sigma-Aldrich) were added to the continuous phase for 3D imaging, and images were taken with a confocal microscope (FV1200, Olympus).

\begin{figure*}
\centering
\includegraphics[width=15.4cm, bb = 0 0 312.240624 70.920142]{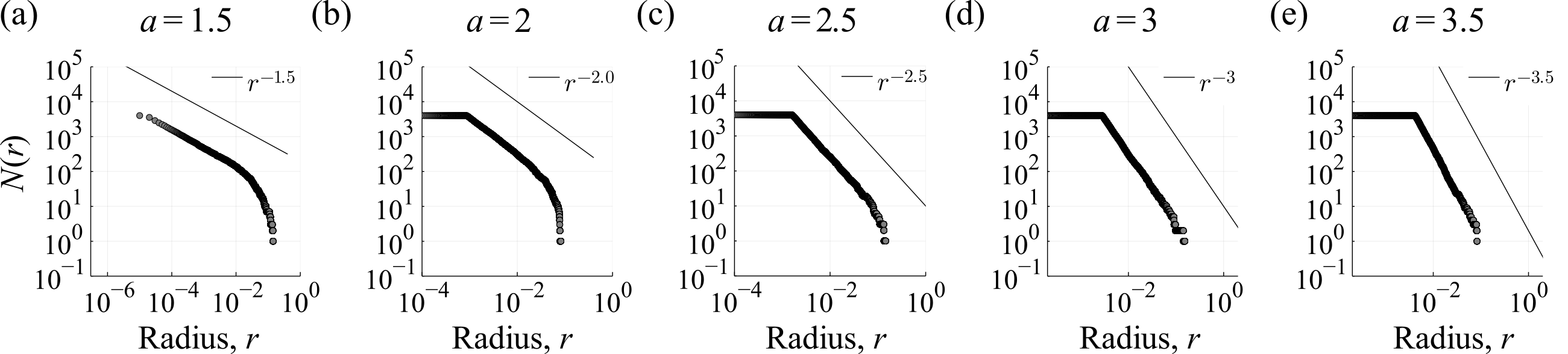}
\caption{Cumulative particle size distributions for $a=1.5, 2, 2.5, 3$, and $3.5$, respectively.}
\label{rplot}
\end{figure*}

Initially, 4,000 circular particles are randomly placed in a square space with periodic boundary conditions in the numerical calculations. The radii of the particles $r$ were randomly set according to a given power size distribution $r^{-a}$ using the inversion method (Supplementary Material, Figure S\ref{rplot}). 
Exponent $a$ was changed to a range $-5<a<10$, keeping the size range $r_{\rm max}/r_{\rm min}=10^2$, where $r_{\rm min}$ and $r_{\rm max}$ are the lower and upper bounds of the distribution, respectively. When calculating the contact number distribution for $a=1.5$ (shown in Fig. S2(a)), and $r_{\rm max}/r_{\rm min}$ was set to $10^5$. This is because when $a$ is small, the $z$-distribution becomes too narrow, and many particles behave as rattlers. The particle size distributions are shown in Figure S2. The cumulative distribution makes the cutoff $r_{\rm max}$ affect regions with large $r$ region, resulting in a downward deviation of the plot.
Increasing $r$ of all particles in the same ratio increased the packing fraction. 
The particles have a repulsive potential $U$:
\begin{equation}
U=\left(D_{ij}-r_{i}-r_{j}\right)^\frac{3}{2}\left(r_i+r_j\right)^\frac{1}{2},
\end{equation}
where $D_{ij}$ is the distance between the $i$-th and $j$-th particles, and $r_{i}$ and $r_{j}$ are the radii of the $i$-th and $j$-th particles, respectively.
The initial placement required 3,200,000 steps to relax the randomly packed pattern using the FIRE algorithm \cite{bitzek2006structural} and the packing fraction was then increased six points over 3,200,000 steps. 
Just after the sum of potential energy exceeded $5\times10^{-4}$, the particles immediately stopped expanding and took 1,600,000 steps to wait for the relaxation of their position.

\bibliographystyle{apsrev4-2}
\bibliography{poly_abrv2}